\documentclass[a4paper,11pt]{article}
\usepackage{amsmath}        
\usepackage{amssymb}        
\usepackage{cite}
\usepackage{cancel}
\usepackage{fancybox}
\usepackage[dvips]{color}
\usepackage{url}
\usepackage{ulem}
\usepackage{pifont}
\usepackage{slashbox}
\usepackage{colortbl}
\usepackage{multicol}
\usepackage{lscape}
\usepackage{colortbl}
\usepackage{multicol}
\usepackage[all,knot]{xy}
\usepackage{longtable}

\makeatletter
 
 \@addtoreset{equation}{section}
\makeatother

\begin{document}

\title{Special Grand Unification}

\author{Naoki Yamatsu
\footnote{Electronic address: yamatsu@cc.kyoto-su.ac.jp}
\\
{\it\small Maskawa Institute for Science and Culture, 
Kyoto Sangyo University, Kyoto 603-8555, Japan}
}
\date{\today}

\maketitle

\begin{abstract}
We discuss new-type grand unified theories based on grand unified groups
broken to their special subgroups as well as their regular subgroups.
In the framework, when we construct four-dimensional (4D) chiral gauge
theories, i.e., the Standard Model (SM), 4D gauge anomaly cancellation
restricts the minimal number of generations of the 4D SM Weyl fermions.
We show that in a six-dimensional (6D) $SU(16)$ gauge theory on 
$M^4\times T^2/\mathbb{Z}_2$ 
one generation of the SM fermions can be embedded into 
a 6D bulk Weyl fermion.
For the model including three chiral generations of the SM fermions,
the 6D and 4D gauge anomalies on the bulk and fixed points are canceled
out without exotic 4D chiral fermions. 
\end{abstract}

\section{Introduction}

One of the most attractive idea to construct unified theories beyond
the standard model (SM) is grand unification since it appeared
\cite{Georgi:1974sy}. 
It is known in e.g., Ref.~\cite{Slansky:1981yr,Yamatsu:2015gut} that the
candidates for grand unified theory (GUT) gauge groups in
four-dimensional (4D) theories are only  
$SU(n) (n\geq 5)$, $SO(4n+2) (n\geq 2)$, and $E_6$ because of 
rank and type of representations. For examples, there are GUTs based on 
$SU(5)$\cite{Georgi:1974sy}, 
$SU(6)$\cite{Inoue:1977qd}, 
$SO(10)$\cite{Fritzsch:1974nn}, 
$SO(14)$\cite{Ida:1980ea},
$SO(18)$\cite{Fujimoto:1981bv},
$E_6$\cite{Gursey:1975ki} gauge groups.
Recently, in Ref.~\cite{Yamatsu:2015gut}, the author showed 
that the candidates for GUT gauge groups in higher dimensional
theories are $SU(n) (n\geq 5)$, $SO(n) (n\geq 9)$, $USp(2n) (n\geq 4)$, 
$E_n (n=6,7,8)$, $F_4$.
The candidates for GUT gauge groups in gauge-Higgs grand unified
theories are $SU(n) (n\geq 6)$, $SO(n) (n\geq 11)$, $USp(2n) (n\geq 5)$, 
$E_n (n=6,7,8)$, $F_4$.
Five-dimensional (5D) gauge theories based on e.g.,
$SU(5)$\cite{Kojima:2011ad,Kojima:2016fvv},
$SU(6)$\cite{Burdman:2002se,Lim:2007jv},
$SU(7)$\cite{Kojima:2017qbt},
$SU(8)$\cite{Kojima:2017qbt},
$SO(10)$\cite{Kim:2002im,Fukuyama:2008pw},
$SO(11)$\cite{Hosotani:2015hoa,Hosotani:2015wmb,Yamatsu:2015rge,Furui:2016owe,Hosotani:2016njs},
$E_6$\cite{Kawamura:2013rj,Kojima:2017qbt} gauge groups
and six-dimensional (6D) ones based on e.g.,
$SO(11)$\cite{Hosotani:2017ghg}
are discussed.

More than half a century ago, E.~Dynkin showed in
Refs.~\cite{Dynkin:1957ek,Dynkin:1957um} that simple Lie algebras has
not only regular subalgebras but also special (or irregular) subalgebras.
All regular subalgebras are obtained by deleting dots from
Dynkin diagrams, while special subalgebras are not. For example, for an
embedding $SU(3)\supset SU(2)\times U(1)$, its 
$SU(2)\times U(1)$ is a maximal regular subalgebra of $SU(3)$, while for
an embedding $SU(3)\supset SU(2)$, its $SU(2)$ is a maximal special
subalgebra. One of the branching rules of the regular subalgebra 
$SU(3)\supset SU(2)\times U(1)$ is 
${\bf 3}=({\bf 2})(1)\oplus({\bf 1})(-2)$, 
where ${\bf 3}$ in the left-hand side stands for an $SU(3)$
representation, and $({\bf 2})(1)$ and ${\bf 1}(-2)$ in the right-hand
side show $SU(2)\times U(1)$ representations. 
(The same rule will be applied below.)
One of the branching rules of the special subalgebra 
$SU(3)\supset SU(2)$ is ${\bf 3}=({\bf 3})$.
For an embedding $G\supset H$, a maximal subalgebra $H$ is a subalgebra
of $G$ if there is no larger subalgebra containing it except $G$ itself.
(See e.g., Refs.~\cite{Cahn:1985wk,Yamatsu:2015gut} in detail.)

In principle, we can construct GUTs based on Lie groups broken to not
only their regular subgroups but also their special subgroups. 
However, at present, there seem to be no GUTs by using special
embeddings, which is referred as special grand unification or special
GUTs. 
In the following, we consider large GUT gauge groups $G$ broken to
smaller GUT groups $H$, called small GUT gauge groups. The groups $H$ 
is regarded as usual GUT groups e.g., $SU(5)$, $SO(10)$, $E_6$.  
When we apply them for a special GUT model building, the rank of its
large GUT gauge groups is usually much larger than small GUT gauge groups.
Even a famous GUT review paper \cite{Slansky:1981yr} written by
R.~Slansky does not contain enough information to construct special
GUTs. Recently, simple Lie algebras with up to rank-15 and $D_{16}$ and
their regular and special subalgebras are given in
Ref.~\cite{Yamatsu:2015gut}, where the information in its tables 
is not enough to survey special GUTs.

We can find candidates for large GUT gauge groups in special GUTs
by using usual methods about branching rules of Lie algebras under their
maximal subalgebras discussed in Refs.~\cite{Cahn:1985wk,Yamatsu:2015gut}.
Under usual requirements for their small GUT gauge groups that have
complex representations and contain the SM gauge group 
$G_{\rm SM}:=SU(3)_C\times SU(2)_L\times U(1)_Y$
the candidates for large GUT gauge groups broken to maximal special
subgroups in 4D theories are e.g.,  
$SU(\frac{n(n+1)(n+2)}{6})\supset SU(n)$, 
$SU(\frac{n(n-1)(n+1)}{3})\supset SU(n)$, 
$SU(\frac{n(n-1)}{2})\supset SU(n)$, 
$SU(\frac{n(n+1)}{2})\supset SU(n)$ $(n\geq 5)$,
$SU(2^{2k})\supset SO(4k+2)$ $(k\geq 2)$ and $SU(27)\supset E_6$.
In five- or higher-dimensional theories, the requirement for the small
GUT gauge groups that contain the SM gauge group $G_{\rm SM}$ leads to 
additional candidates e.g., 
$SO(n^2-1)\supset SU(n)$ $(n\geq 5)$,
$USp(2^{4\ell+1})\supset SO(8\ell+4)$ $(\ell\geq 1)$,
$USp(56)\supset E_7$,
$SO(248)\supset E_8$,
$SO(26)\supset F_4$,
where almost the above pairs are explicitly shown by E.~Dynkin in
Ref.~\cite{Dynkin:1957ek}. 

To construct unified theories beyond the SM based on a GUT gauge 
group $G_{\rm GUT}\supset G_{\rm SM}$, 
we must take into account for not only how to realize the SM gauge 
group at a vacuum but also how to embed the SM matter content into 
the GUT matter content.
To do so, we need the information about the branching rules of
Lie algebras under their subalgebras.
It is known that their branching rules can be calculated by using their
corresponding projection matrices introduced in Ref.~\cite{Mckay:1977}. 
Many examples of the branching rules of simple Lie algebras with up to
rank 8 under their regular and special semi-simple subalgebras are
listed in Ref.~\cite{McKay:1981}. Recently, more examples up to rank-15
and $SO(32)$ including non-semi-simple subalgebras are
given in Ref.~\cite{Yamatsu:2015gut}.
Other cases can be calculated by using appropriate computer programs,
such as Susyno program \cite{Fonseca:2011sy} and LieART
\cite{Feger:2012bs}, 
and appropriate projection matrices, which can be calculated by using 
a usual weight diagram method discussed in 
Refs.~\cite{Cahn:1985wk,Yamatsu:2015gut}.

It is also known that one generation of the five SM left-handed Weyl
fermions, which consist of a quark doublet, up- and down-type
quarks, a lepton doublet and a charged lepton, can be embedded into an
$SU(5)$ reducible representation 
$\left({\bf 10}\oplus{\bf \overline{5}}\right)$, where we will omit
left-handed if it is not needed. 
One generation of the SM Weyl fermions and vectorlike
fermions can be embedded into an $SU(n)$ $(n\geq 6)$ reducible
representation 
$\left({\bf \frac{n(n-1)}{2}}\oplus(n-4){\bf \overline{n}}\right)$.
(Note that their 4D $SU(n)$ gauge anomaly coefficients from the 4D Weyl 
fermions is zero. It can be checked by using tables in
Ref.~\cite{Yamatsu:2015gut}.) 
One generation of the SM fermions and a SM singlet fermion can be
embedded into an $SO(10)$ spinor representation ${\bf 16}$.
For the other candidates for rank-4 and -5 GUT gauge groups 
$SO(9)$, $USp(8)$, $F_4$, $SO(11)$, and $USp(10)$,
it is examined which representations of them can contain the SM fermions
in Ref.~\cite{Yamatsu:2015gut}.

There are several good features of special GUTs. First, we can
eliminate almost all unnecessary $U(1)$s even if we consider very large
GUT gauge groups. For example, the rank of a gauge group $SO(32)$ is
sixteen. We have the SM gauge group and twelve extra $U(1)$s if we use
only regular embeddings, while we have only the SM gauge group and two
extra $U(1)$s if we use a special embedding  
$SU(16)\supset SO(10)$. Second, by using only regular embeddings, we
cannot embed the representations of the SM fermions into an
$SO(32)$ vector representation ${\bf 32}$, while by using regular and
special embeddings 
$SO(32)\supset SU(16)\times U(1)\supset SO(10)\times U(1)$,
the representations of the SM fermions can be embedded into an
$SO(32)$ vector representation because 
an $SO(32)$ vector representation ${\bf 32}$ 
is decomposed into $SO(10)$ spinor representations ${\bf 16}$ and 
${\bf \overline{16}}$ \cite{Yamatsu:2015gut}.
Third, asymptotic freedom may be realized in special GUTs. This is
because the larger contribution comes from large GUT gauge fields
while the smaller contribution comes from the GUT fermions.

In this paper, we discuss special GUTs based on a large GUT gauge group
$G$ broken to its maximal special subgroup $H(\subset G)$, where $H$ is
regarded as a usual GUT gauge group.
In 4D framework, to realize a 4D chiral gauge theory, 
the large GUT groups $G$ and their maximal special subgroups $H$ must
contain complex representations. In the following discussion, we focus
on $G=SU(n)$ with less than rank-$30$ and $H=SU(k), SO(4k+2), E_6$.
Also, the special groups $H$ must have rank four or greater because
the rank of the SM gauge group $G_{\rm SM}$ is four.
The conditions are satisfied by the following nine pairs of Lie
groups and their maximal special subgroups:
$SU(\frac{n(n-1)}{2})\supset SU(n)$ $(n=5,6,7,8)$;
$SU(\frac{n(n+1)}{2})\supset SU(n)$ $(n=5,6,7)$;
$SU(16)\supset SO(10)$; and 
$SU(27)\supset E_6$.

We find the following facts from Tables in Ref.~\cite{Yamatsu:2015gut}
and its calculating methods.
We consider 4D $SU(n)$ $(n\leq 30)$ special GUTs whose fermions are a 4D
Weyl fermion in an $SU(n)$ reducible representation  
$\left((n-4){\bf n}\oplus{\bf \overline{\frac{n(n-1)}{2}}}\right)$,
where the representation is 4D anomaly-free. 
Only for the $SU(16)\supset SO(10)$ and
$SU(27)\supset E_6$ cases, their matter contents contain the SM 
chiral fermions.
For the $SU(16)\supset SO(10)$ case, the fermion is a 4D 
$SU(16)$ $\left(12\times{\bf 16}\oplus{\bf \overline{120}}\right)$
Weyl fermion, which stands for a 4D left-handed Weyl fermion in
an $SU(16)$ reducible representation 
$\left(12\times{\bf 16}\oplus{\bf \overline{120}}\right)$,
where the similar notation is used below.
As we will see later, for 4D framework there are twelve generations of
the SM Weyl fermions but fortunately there is no exotic non-SM chiral
fermions due to a special property of the $SU(16)$ complex
representation ${\bf \overline{120}}$. 
For the $SU(27)\supset E_6$ case, its matter content contains
non-SM chiral fermions.

The main purpose of this paper is to show that 
in a 6D $SU(16)$ special GUT on $M^4\times T^2/\mathbb{Z}_2$
we can realize three generations of the 4D SM Weyl fermions 
from twelve 6D $SU(16)$ ${\bf 16}$ bulk Weyl fermions.

This paper is organized as follows. In Sec.~\ref{Sec:Special-GUT},
we first discuss basic properties of a special embedding 
$SU(16)\supset SO(10)$ in 4D framework. After that, by using the special
embedding, we construct a 6D $SU(16)$ special GUT on 
$M^4\times T^2/\mathbb{Z}_2$.
Section~\ref{Sec:Summary-discussion} is devoted to a summary and
discussion.

\section{Special grand unification}
\label{Sec:Special-GUT}

Before we start to construct 6D special GUTs, we examine how to embed
the SM Weyl fermions into $SU(16)$ GUT multiplets in purely 4D framework.
For a special embedding $SU(16)\supset SO(10)$, 
an $SU(16)$ fundamental representation ${\bf 16}$
$({\bf \overline{16}})$ is identified as 
an $SO(10)$ spinor representation ${\bf 16}$
$({\bf \overline{16}})$
\cite{Yamatsu:2015gut}:
\begin{align}
{\bf 16}={\bf 16}\ \ \
({\bf \overline{16}}={\bf \overline{16}}).
\end{align}
Further, the $SO(10)$ spinor representation ${\bf 16}$ is decomposed as
usual into 
$G_{\rm SM}\times U(1)_X=SU(3)_C\times SU(2)_L\times U(1)_Y\times U(1)_X$
representations 
\begin{align}
{\bf 16}=&
({\bf 3,2})(-1)(1)
\oplus({\bf \overline{3},1})(-2)(-3)
\oplus({\bf \overline{3},1})(4)(1)\nonumber\\
&\oplus({\bf 1,2})(3)(-3)
\oplus({\bf 1,1})(-6)(1)
\oplus({\bf 1,1})(0)(5),
\end{align}
where the convention of their $U(1)$ normalization is the same as one in
Ref.~\cite{Yamatsu:2015gut}.
Obviously, we can identify an $SU(16)$ ${\bf 16}$ fermion as one
generation of the SM fermions plus a right-handed neutrino. 
The 4D anomaly coefficient of an $SU(16)$ ${\bf 16}$ fermion is
non-zero, while the 4D anomaly coefficient of any $SO(10)$ fermion is 
zero.

For $SU(16)$ models instead of $SO(10)$ ones, we cannot
construct 4D anomaly-free theories with a chiral matter content 
by using only 4D $SU(16)$ ${\bf 16}$ and/or ${\bf \overline{16}}$ Weyl
fermions. We need to introduce an additional Weyl fermion to cancel out 4D
$SU(16)$ gauge anomaly. A primary candidate is a 4D $SU(16)$
 ${\bf \overline{120}}$ Weyl fermion, where 
${\bf \overline{120}}$ is the second lowest dimensional complex
representation of $SU(16)$ and has the second smallest value of the 4D
anomaly coefficient in $SU(16)$ complex representations with at
least up to $10^{8}$ dimension. Its anomaly coefficient of $SU(16)$ is
$-12$.
By using $SU(16)$ complex representations ${\bf 16}$ and 
${\bf \overline{120}}$, we can realize a 4D anomaly-free chiral gauge
theory whose fermion content is a 4D 
$SU(16)$ $\left(12\times{\bf 16}\oplus{\bf \overline{120}}\right)$
Weyl fermion.

From Tables in Ref.~\cite{Yamatsu:2015gut} for the branching rules of
$SU(16)\supset SO(10)$, we find that an $SU(16)$ complex representation
${\bf \overline{120}}$ (${\bf 120}$) is just identified as an $SO(10)$
real representation ${\bf 120}$:
\begin{align}
{\bf \overline{120}}={\bf 120}\ \ \
({\bf 120}={\bf 120}).
\end{align}
We consider a 4D $SU(16)$ special GUT whose fermion content is 
an $SU(16)$ $\left(12\times{\bf 16}\oplus{\bf \overline{120}}\right)$
Weyl fermion.
When $SU(16)$ is broken to $SO(10)$ via a non-vanishing VEV of a GUT
breaking Higgs scalar boson, there are twelve generations of $SO(10)$
${\bf 16}$ Weyl fermions. 
One may wonder what kind of GUT breaking Higgs can reduce $SU(16)$ to
$SO(10)$, properly. The lowest dimensional $SU(16)$ representation is 
${\bf 5440}$ $({\bf \overline{5440}})$
that contains singlet under the $SO(10)$ special subgroup. Its $SO(10)$
decomposition is shown in Ref.~\cite{Yamatsu:2015gut}:
\begin{align}
{\bf 5440}=
{\bf 4125}
\oplus{\bf \overline{1050}}
\oplus{\bf 210}
\oplus{\bf 54}
\oplus{\bf 1}\ \ \
({\bf \overline{5440}}=
{\bf 4125}
\oplus{\bf 1050}
\oplus{\bf 210}
\oplus{\bf 54}
\oplus{\bf 1}).
\end{align}

An additional question is how to break $SU(16)$ to $G_{\rm SM}$. One way
of achieving it is to use several GUT breaking Higgses, where we assume
their proper components get non-vanishing VEVs. One example is
to introduce three $SU(16)$ ${\bf 5440}$, ${\bf 255}$, ${\bf 16}$
scalar fields.
In this case, the non-vanishing VEV of the $SU(16)$ ${\bf 5440}$ scalar
field is responsible for breaking $SU(16)$ to $SO(10)$;
the additional VEV of the $SU(16)$ ${\bf 16}$ scalar
breaks $SU(16)\supset SO(10)$ to $SU(5)$;
the last VEV of the $SU(16)$ ${\bf 255}$ scalar
reduces $SU(16)\supset SO(10)\supset SU(5)$ to $G_{\rm SM}$, where 
the $SU(16)$ adjoint representation ${\bf 255}$ is decomposed into
$SO(10)$ representations 
\begin{align}
{\bf 255}={\bf 210}\oplus{\bf 45}.
\end{align}
It may be possible to reduce the number of the scalar
fields e.g., by using orbifold BCs in extra dimension 
\cite{Kawamura:1999nj,Kawamura:2000ev}
or the Hosotani mechanism
\cite{Hosotani:1983xw,Hosotani:1988bm}. 

We start to discuss an $SU(16)$ special GUT on 6D orbifold
spacetime $M^4\times T^2/\mathbb{Z}_2$ with 
the Randall-Sundrum (RS) type metric
\cite{Randall:1999ee,Hosotani:2017krs,Hosotani:2017ghg} 
given by 
\begin{align}
ds^2=e^{-2\sigma(y)}(\eta_{\mu\nu}dx^{\mu}dx^{\nu}+dv^2)+dy^2,
\end{align}
where $\eta_{\mu\nu}=\mbox{diag}(-1,+1,+1,+1)$, 
$y$ is the fifth coordinate of RS warped space,
$v$ is the sixth coordinate of $S^1$ ($v\sim v+2\pi R_6$) and 
$\sigma(y)=\sigma(-y)=\sigma(y+2\pi R_5)$,
$\sigma(y)=k|y|$ for $|y|\leq \pi R_5$.
The space has four fixed points on $T^2/\mathbb{Z}_2$ 
at $(y_0,v_0)=(0,0)$, $(y_1,v_1)=(\pi R_5,0)$, 
$(y_2,y_2)=(0,\pi R_6)$, and $(y_3,v_3)=(\pi R_5,\pi R_6)$.
The $\mathbb{Z}_2$ parity reflection around each fixed point is given by 
\begin{align}
P_j:\ (x_\mu,y_j+y,v_j+v)\ \to\ (x_\mu,y_j-y,v_j-v),
\end{align}
where $j=0,1,2,3$. Note that only three of them is independent because
of $P_3=P_1P_0P_2=P_2P_0P_1$.  
Also, 5th and 6th dimensional translation 
$U_5: (x_\mu,y,v)\to(x_\mu,y+2\pi R_5,v)$ 
and $U_6: (x_\mu,y,v)\to(x_\mu,y,v+2\pi R_6)$  
can be described as $U_5=P_1P_0$ and $U_6=P_2P_0$, respectively.

The orbifold BCs of a 6D $SU(16)$ ${\bf 16}$ positive or negative Weyl
bulk fermion can be written by 
\begin{align}
\Psi_{{\bf 16}\pm}(x,y_j-y,v_j-v)&=
\eta_{j}(-i\Gamma^5\Gamma^6)
P_{j{\bf 16}}
\Psi_{{\bf 16}\pm}
(x,y_j+y,v_j+v),
\label{Eq:BC-SU(16)-fermion-16}
\end{align}
where the subscript of $\Psi$ $\pm$ stands for 6D chirality,
$\eta_{j}$ is a positive or negative sign,
$\Gamma^M$ $(M=1,2,\cdots,7)$ is a 6D gamma matrix, and
$P_{j{\bf 16}}$ is a projection matrix acting on the $SU(16)$
representation ${\bf 16}$.
In the following discussion, we take $P_{j{\bf 16}}=1$,
but still we need to choose $\eta_j$, where  $\prod_{j=0}^{3}\eta_j=1$
must be satisfied because of consistency of orbifold BCs
\cite{Hosotani:2017ghg}.

We introduce three copies of the sets of two 6D $SU(16)$ ${\bf 16}$
positive and negative Weyl fermion pairs. They must have appropriate
orbifold BCs to realize 6D bulk and 4D brane gauge anomaly-free, and 
three chiral generations of quarks and leptons. More explicitly, each
set of 6D Weyl fermions consists of a 6D $SU(16)$ ${\bf 16}$ positive
Weyl fermion with orbifold BCs
$(\eta_{0},\eta_{1},\eta_{2},\eta_{3})=(-,-,-,-)$;
a 6D positive one with orbifold BCs $(-,-,+,+)$;
a 6D negative one with orbifold BCs $(-,+,+,-)$; and
a 6D negative one with orbifold BCs $(-,+,-,+)$.
In this case, only the first 6D positive Weyl fermion with the orbifold
BCs $(-,-,-,-)$ has zero modes for
its 4D (left-handed) Weyl fermion components because
its 4D left-handed Weyl fermion components have Neumann BCs
at all the fixed points, 
while its 4D right-handed Weyl fermion components have Dirichlet BCs
at all the fixed points.
For the other three fermions, all their components have two
Neumann and two Dirichlet BCs at four fixed points $(y_j,v_j)$.

From the above discussion, we find that in 4D framework the 4D
anomaly-free $SU(16)$ chiral gauge theories cannot realize the SM matter
content because the 4D $SU(16)$ anomaly cancellation does not allow three
chiral generation of quarks and leptons.
The 4D anomaly situation is exactly the same as each 4D
fixed point in 6D theories with two dimensional orbifold extra dimension
$T^2/\mathbb{Z}_2$. However, as we will see below, even when effectively
twelve 4D $SU(16)$ ${\bf 16}$ Weyl fermions from six 6D
$SU(16)$ ${\bf 16}$ positive and negative Weyl fermions exist at an 4D
fixed point, three 6D positive Weyl fermions have zero modes and their
zero modes are only 4D $SU(16)$ ${\bf 16}$ Weyl fermion.
When we take into account appropriate symmetry breaking effects, the
zero modes become three generations of the SM fermions. 

Here we check the contribution to 6D bulk and 4D brane anomalies from
the above four 6D Weyl fermion set. 
Obviously, the fermion set does not contribute to 6D gauge anomaly
because the number of 6D $SU(16)$ ${\bf 16}$ positive and negative Weyl
fermions are the same. In other words, the fermion set forms 6D
vectorlike fermions in the bulk. 
We calculate 4D gauge anomaly numbers at four fixed points
$(y_j,v_j) (j=0,1,2,3)$ by using 4D $SU(16)$ anomaly coefficients listed
in Ref.~\cite{Yamatsu:2015gut}, where for our convention the anomaly
coefficient of a 4D $SU(16)$ ${\bf 16}$ Weyl fermion is $+1$. 
The above first 6D positive Weyl fermion contributes $+1$ 
to the $SU(16)$ anomaly number 
for all the four fixed points $(y_j,v_j) (j=0,1,2,3)$;
the second 6D positive Weyl fermion contributes 
$+1$ for the fixed points $(y_0,v_0)$ and $(y_1,v_1)$ and 
$-1$ for the fixed points $(y_2,v_2)$ and $(y_3,v_3)$;
the third 6D negative Weyl fermion contributes 
$+1$ for the fixed points $(y_0,v_0)$ and $(y_3,v_3)$ and 
$-1$ for the fixed points $(y_1,v_1)$ and $(y_2,v_2)$;
the last 6D negative Weyl fermion contributes 
$+1$ for the fixed points $(y_0,v_0)$ and $(y_2,v_2)$ and 
$-1$ for the fixed points $(y_1,v_1)$ and $(y_3,v_3)$.
Therefore, one set of the four 6D $SU(16)$ ${\bf 16}$ Weyl fermions
contributes $+4$ for the fixed point $(y_0,v_0)$ and 
$0$ for the other three fixed points $(y_1,v_1)$, $(y_2,v_2)$, and 
$(y_3,v_3)$.
Since the SM has three chiral generations of quarks and leptons, we need
to introduce three sets of the four 6D Weyl fermions.
Its total anomaly number from three set of the four 6D Weyl fermions 
are $+12$ for the fixed point $(y_0,v_0)$ and 
$0$ for the other fixed points $(y_1,v_1)$, $(y_2,v_2)$, and $(y_3,v_3)$.

To cancel the 4D $SU(16)$ anomaly on all the fixed points, we need to
introduce at least one 4D Weyl fermion of an $SU(16)$ non-anomaly-free
complex representation on the fixed point $(y_0,v_0)$. As we saw before, 
a 4D $SU(16)$ $\overline{\bf 120}$ Weyl fermion 
contributes $-12$ to the $SU(16)$ anomaly number.
Therefore, by introducing a 4D $SU(16)$ $\overline{\bf 120}$
Weyl brane fermion at the fixed point $(y_0,v_0)$, all the 4D $SU(16)$
anomalies at the fixed points are canceled out. 

In the 6D framework, we introduce three 5D $SU(16)$ ${\bf 5440}$, 
${\bf 255}$, ${\bf 16}$ brane scalar fields $\Phi_{\bf 5440}$, 
$\Phi_{\bf 255}$, $\Phi_{\bf 16}$ on the 5D brane ($y=0$).
Their orbifold BCs are given by 
\begin{align}
\Phi_{\bf x}(x,v_\ell-v)=
\eta_{\ell{\bf x}}P_{\ell{\bf x}}\Phi_{\bf x}(x,v_\ell+v),
\end{align}
where $\ell=0,2$, ${\bf x}={\bf 5440},{\bf 255},{\bf 16}$,
$\eta_{\ell{\bf x}}$ is a positive or negative sign, and
$P_{\ell{\bf x}}$ is a projection matrix. In our orbifold BCs, 
$P_{\ell{\bf x}}=1$.
The scalar fields are responsible for breaking $SU(16)$ to $G_{\rm SM}$.
When we take $\eta_{\ell{\bf x}}=1$, they have zero modes
and can get nonvanishing VEVs. We assume that the nonvanishing VEVs
of the scalars break $SU(16)$ to $G_{\rm SM}$.

\begin{table}[t]
\begin{center}
\begin{tabular}{cccccc}\hline
\rowcolor[gray]{0.8}
6D Bulk field&
$A_M$&$\Psi_{{\bf 16}+}^{(a)}$&$\Psi_{{\bf 16}+}^{(b)}$
&$\Psi_{{\bf 16}-}^{(c)}$&$\Psi_{{\bf 16}-}^{(d)}$
\\\hline
$SU(16)$ &${\bf 255}$&${\bf 16}$&${\bf 16}$&${\bf 16}$&${\bf 16}$\\
$SO(5,1)$&${\bf 6}$
&${\bf 4}_+$&${\bf 4}_+$&${\bf 4}_-$&${\bf 4}_-$\\
Orbifold BC&
&$\left(
\begin{array}{cc}
-&-\\
-&-\\
\end{array}
\right)$
&$\left(
\begin{array}{cc}
+&+\\
-&-\\
\end{array}
\right)$
&$\left(
\begin{array}{cc}
+&-\\
-&+\\
\end{array}
\right)$
&$\left(
\begin{array}{cc}
-&+\\
-&+\\
\end{array}
\right)$\\
\hline
\end{tabular}\\[0.5em]
\begin{tabular}{cccc}\hline
\rowcolor[gray]{0.8}
5D Brane field
&$\Phi_{\bf 5440}$&$\Phi_{\bf 255}$&$\Phi_{\bf 16}$\\
\hline
$SU(16)$          &${\bf 5440}$&${\bf 255}$&${\bf 16}$\\
$SO(4,1)$&{\bf 1}&{\bf 1}&{\bf 1}\\
Orbifold BC
&$\left(
\begin{array}{c}
+\\
+\\
\end{array}
\right)$
&$\left(
\begin{array}{c}
+\\
+\\
\end{array}
\right)$
&$\left(
\begin{array}{c}
+\\
+\\
\end{array}
\right)$\\
Spacetime  &$y=0$&$y=0$&$y=0$\\
\hline
\end{tabular}
\hspace{1em}
\begin{tabular}{cc}\hline
\rowcolor[gray]{0.8}
4D Brane field&$\psi_{\overline{\bf 120}}$\\
\hline
$SU(16)$          &$\overline{\bf 120}$\\
$SL(2,\mathbb{C})$&$(1/2,0)$\\
Spacetime $(y,v)$ &$(0,0)$\\
\hline
\end{tabular}
\end{center}
\caption{The table shows the matter content in
the $SU(16)$ special GUT on $M^4\times T^2/\mathbb{Z}_2$.
The representations of $SU(16)$ and 6D, 5D, 4D Lorentz group, 
the orbifold BCs of 6D bulk fields and 5D brane fields, and 
the spacetime location of 5D and 4D brane fields
are shown.
\label{tab:SU16-SO10-matter-content-6D}}
\end{table}

To summarize, the matter content in the $SU(16)$ special GUT 
consists of 
a 6D $SU(16)$ bulk gauge boson $A_{M}$;
three 6D $SU(16)$ ${\bf 16}$ positive Weyl fermions with the orbifold BCs
$(\eta_{0},\eta_{1},\eta_{2},\eta_{3})=(-,-,-,-)$ 
$\Psi_{{\bf 16}+}^{(a)}$ $(a=1,2,3)$;
three 6D positive one with $(-,-,+,+)$
$\Psi_{{\bf 16}+}^{(b)}$ $(b=1,2,3)$;
three 6D negative one with $(-,+,+,-)$
$\Psi_{{\bf 16}-}^{(c)}$ $(c=1,2,3)$; 
three 6D negative one with $(-,+,-,+)$
$\Psi_{{\bf 16}-}^{(d)}$ $(d=1,2,3)$;
three 5D $SU(16)$ ${\bf 5440}$, ${\bf 255}$, ${\bf 16}$ brane scalar
bosons at $y=0$
$\Phi_{\bf 5440}$, $\Phi_{\bf 255}$, $\Phi_{\bf 16}$;
one 4D $SU(16)$ ${\bf \overline{120}}$ Weyl brane fermion at
the fixed point $(y_0,v_0)=(0,0)$ $\psi_{\bf \overline{120}}$.
The matter content of the $SU(16)$ special GUT is summarized in 
Table~\ref{tab:SU16-SO10-matter-content-6D}.

\section{Summary and discussion}
\label{Sec:Summary-discussion}

In this paper, we constructed $SU(16)$ special GUTs by using a special
embedding $SU(16)\supset SO(10)$. In this framework, we found that 
the 4D $SU(16)$ ${\bf 16}$ Weyl fermion can be identified with
one generation of quarks and leptons;
the 4D $SU(16)$ gauge anomaly on the fixed points restricts the minimal
number of generations; 
we cannot construct $SU(16)$ special GUTs with three generations of
quarks and leptons in 4D framework, while 
we can construct an $SU(16)$ special GUT in 6D framework;
also exotic chiral fermions do not exist due to a special feature of
the $SU(16)$ complex representation ${\bf \overline{120}}$ once we take
into account the symmetry breaking of the large GUT gauge group $SU(16)$
to the small GUT gauge group $SO(10)$. 

We comment on the SM fermion masses in the 6D $SU(16)$ special GUT. 
Since one generation of the SM fermions is unified into a 6D $SU(16)$
${\bf 16}$ Weyl fermion, the masses of quarks and leptons for each
generation are degenerate without $SU(16)$ breaking effects. 
We introduced a 5D $SU(16)$ ${\bf 16}$ brane fermion and a 4D $SU(16)$
${\bf \overline{120}}$ brane fermion at a fixed point $(y_0,v_0)=(0,0)$.
The $SO(10)$ ${\bf 16}$ and ${\bf 120}$ representations are decomposed
into $SU(5)\times U(1)$ representations
\begin{align}
{\bf 16}&=({\bf 10})(1)\oplus({\bf \overline{5}})(-3)\oplus({\bf 1})(5),\\
{\bf 120}&=({\bf \overline{45}})(2)
\oplus({\bf 45})(-2)
\oplus({\bf 5})(-2)
\oplus({\bf 10})(6)
\oplus({\bf \overline{10}})(-6)
\oplus({\bf \overline{5}})(2).
\end{align}
Since an $SO(10)$ tensor product 
${\bf 16}\otimes{\bf 16}\otimes{\bf 120}$ contains singlet,
the 6D $SU(16)$ ${\bf 16}$ bulk and 4D $SU(16)$ ${\bf \overline{120}}$
brane fermions can be mixed via the VEV of the 5D $SU(16)$ ${\bf 16}$
brane scalar once its corresponding brane interaction term is generated.
The effective mass term divides up-type quark mass with down-type quark
and charged lepton masses. Next, we introduced a 5D $SU(16)$ ${\bf 255}$
brane scalar field on the brane $y=0$ and its VEV is responsible for
breaking $SU(5)(\subset SU(16))$ to $G_{\rm SM}$. The VEV of the brane
scalar field can mix three 6D $SU(16)$ ${\bf 16}$ bulk Weyl fermions 
containing zero modes with the other nine 6D $SU(16)$ ${\bf 16}$ ones.
Therefore, the $SU(16)$ model seems to realize the SM fermion masses,
but seems to give us no prediction about quark and lepton masses. 
We will leave the detail analysis in future studies.

In the 6D $SU(16)$ special GUT, we use the fact that an $SU(16)$ complex
representation ${\bf \overline{120}}$ is identified with an $SO(10)$ real
representation ${\bf 120}$ for the special embedding 
$SU(16)\supset SO(10)$. 
It is important to eliminate non-SM chiral fermions at low-energy
physics. This type of representations, which is complex under a Lie
group but is real under its maximal special subgroup, 
cannot be found in the other special embeddings $G\supset H$ with up to
the rank-$30$ of $G$ except $G=SU(27)\supset H=E_6$.
(It is checked at least for up to $10^4$ dimensional representations of 
$SU(10)\supset SU(5)$ and $SU(15)\supset SU(5)$;
up to $10^5$ dimensional representations of 
$SU(15)\supset SU(6)$, 
$SU(21)\supset SU(6)$, $SU(21)\supset SU(7)$,
$SU(28)\supset SU(7)$, $SU(28)\supset SU(8)$.)
For a special embedding $SU(27)\supset E_6$, an $SU(27)$
complex representation ${\bf \overline{2925}}$ $({\bf 2925})$
becomes an $E_6$ real representation ${\bf 2925}$:
\begin{align}
{\bf \overline{2925}}={\bf 2925}\ \ \ ({\bf 2925}={\bf 2925}).
\end{align}
Many branching rules of $SU(10)\supset SU(5)$, 
$SU(15)\supset SU(5)$, and $SU(15)\supset SU(6)$ are listed 
in Ref.~\cite{Yamatsu:2015gut};
for  branching rules of
$SU(21)\supset SU(6)$, $SU(21)\supset SU(7)$,
$SU(28)\supset SU(7)$, $SU(28)\supset SU(8)$,
$SU(27)\supset E_6$,
there are no references, but their branching rules can be calculated by
using methods in Ref.~\cite{Yamatsu:2015gut}. 

We consider a special GUT based on a small GUT gauge group $E_6$.
As the same as $SO(10)$ gauge theories, there is no 4D $E_6$ gauge
anomaly, while there can be 4D $SU(27)$ gauge anomaly in 4D chiral
theories. We have to check how to cancel out the 4D $SU(27)$ gauge
anomaly. When we take the 4D anomaly coefficient of a 4D $SU(27)$ 
${\bf 27}$ Weyl fermion as $+1$, 
the 4D anomaly coefficient of a 4D $SU(27)$ ${\bf \overline{2925}}$
Weyl fermion is $-252$, 
which can be calculated by using the general formula of 4D anomaly
coefficients of $SU(n)$ representations discussed in
Refs.~\cite{Banks1976,Yamatsu:2015gut}. 
Thus, if we consider a model whose fermion matter content is a 4D
$SU(27)$ $\left(252\times {\bf 27}\oplus{\bf \overline{2925}}\right)$
Weyl fermion, its matter content does not contain exotic non-SM chiral
fermions. However, there are $252$ generations of the SM Weyl fermions.
It seems to be impossible to realize three chiral generations of quarks
and leptons. 

In our 6D $SU(16)$ special GUT, we assumed that 5D brane scalar
fields on the 5D brane on 5th dimensional space $y=0$ are responsible
for breaking the large GUT gauge group $SU(16)$ to $G_{\rm SM}$.
However, a part of its symmetry breaking may be replaced by
orbifold BCs and/or the Hosotani mechanism breaking. In the above
discussion, we took the projection matrix $P_{j{\bf 16}}$ $(j=0,1,2,3)$
for the $SU(16)$ representation ${\bf 16}$ as an identity matrix, but we
can choose different ways. 
For example, when we take  
$P_{0{\bf 16}}=P_{1{\bf 16}}=I$ and 
$P_{2{\bf 16}}=P_{3{\bf 16}}=\mbox{diag}(I_{15},-1)$,
the orbifold BCs break $SU(16)$ to $SU(15)\times U(1)$, and we assume
that the non-vanishing VEV of the $SU(16)$ ${\bf 5440}$ GUT Higgs 
breaks the large GUT gauge group $SU(16)$ to the small GUT gauge group
$SO(10)$. The two symmetry breaking combination leaves only $SU(5)$,
where it can be recognized by using the correspondence of the root
vectors between $SU(16)$ and its special subgroup $SO(10)$.
Also, the $SU(16)$ adjoint representation ${\bf 255}$ is decomposed into 
$SO(10)$ representations ${\bf 210}$ and ${\bf 45}$.
The $SO(10)$ representation ${\bf 210}$ includes not only the $SU(5)$
adjoint representation ${\bf 24}$ and but also all the $SU(5)$
fundamental representations  
${\bf 5}$, ${\bf 10}$, ${\bf \overline{10}}$, and ${\bf \overline{5}}$.
It can be used as GUT or the electro-weak symmetry breaking
Higgses in gauge-Higgs GUT scenarios.

The use of the special embedding $SU(16)\supset SO(10)$ may be
interesting for string inspired GUT model builders because one of the
regular embeddings of $SO(32)$ is $SU(16)\times U(1)$. The special
embedding may be incorporated with model buildings in an $SO(32)$
heterotic string theory. For one of the amazing features, $SO(10)$
spinor representations ${\bf 16}$ and  ${\bf \overline{16}}$ are
embedded into an $SO(32)$ vector representation ${\bf 32}$. 

Another motivation may come from string GUT model buildings to realize 
higher-level gauge groups by lower ``costs'' compared with e.g.,
so-called diagonal embeddings
\cite{Dienes:1996yh,Dienes:1996du,Kojima:2011ad}.
For example, by using a diagonal embedding, we need four
$SO(10)_{1}$s to obtain $SO(10)_4$, 
while by using an special embedding $SU(16)_{1}\supset SO(10)_{4}$,
we need one $SU(16)_{1}$,
where the subscript is the index of embedding, which stands for the
ratio of the second order Dynkin indices of corresponding
representations between a Lie algebra and its subalgebra.
The rank in the former is twenty, while one in the latter is
fifteen \cite{Dienes:1996yh}.

\section*{Acknowledgments}

The author would like to thank Yutaka Hosotani,  Kentaro Kojima,
Taichiro Kugo, and Kenji Nishiwaki for valuable comments.

\bibliographystyle{utphys} 
\bibliography{../../arxiv/reference}

\providecommand{\href}[2]{#2}\begingroup\raggedright\begin{thebibliography}{10}

\bibitem{Georgi:1974sy}
H.~Georgi and S.~L. Glashow, ``{Unity of All Elementary Particle Forces},''
\href{http://dx.doi.org/10.1103/PhysRevLett.32.438}{{\em Phys. Rev. Lett.}
  {\bfseries 32} (1974) 438--441}.
%%CITATION = PRLTA,32,438;%%.

\bibitem{Slansky:1981yr}
R.~Slansky, ``{Group Theory for Unified Model Building},''
\href{http://dx.doi.org/10.1016/0370-1573(81)90092-2}{{\em Phys. Rept.}
  {\bfseries 79} (1981) 1--128}.
%%CITATION = PRPLC,79,1;%%.

\bibitem{Yamatsu:2015gut}
N.~Yamatsu, ``{Finite-Dimensional Lie Algebras and Their Representations for
  Unified Model Building},''
\href{http://arxiv.org/abs/1511.08771}{{\ttfamily arXiv:1511.08771 [hep-ph]}}.
%%CITATION = ARXIV:1511.08771;%%.

\bibitem{Inoue:1977qd}
K.~Inoue, A.~Kakuto, and Y.~Nakano, ``{Unification of the Lepton-Quark World by
  the Gauge Group SU(6)},''
\href{http://dx.doi.org/10.1143/PTP.58.630}{{\em Prog.Theor.Phys.} {\bfseries
  58} (1977) 630}.
%%CITATION = PTPKA,58,630;%%.

\bibitem{Fritzsch:1974nn}
H.~Fritzsch and P.~Minkowski, ``{Unified Interactions of Leptons and
  Hadrons},''
\href{http://dx.doi.org/10.1016/0003-4916(75)90211-0}{{\em Ann. Phys.}
  {\bfseries 93} (1975) 193--266}.
%%CITATION = APNYA,93,193;%%.

\bibitem{Ida:1980ea}
M.~Ida, Y.~Kayama, and T.~Kitazoe, ``{Inclusion of Generations in SO(14)},''
\href{http://dx.doi.org/10.1143/PTP.64.1745}{{\em Prog. Theor. Phys.}
  {\bfseries 64} (1980) 1745}.
%%CITATION = PTPKA,64,1745;%%.

\bibitem{Fujimoto:1981bv}
Y.~Fujimoto, ``{SO(18) Unification},''
\href{http://dx.doi.org/10.1103/PhysRevD.26.3183}{{\em Phys. Rev.} {\bfseries
  D26} (1982) 3183}.
%%CITATION = PHRVA,D26,3183;%%.

\bibitem{Gursey:1975ki}
F.~Gursey, P.~Ramond, and P.~Sikivie, ``{A Universal Gauge Theory Model Based
  on $E_6$},''
\href{http://dx.doi.org/10.1016/0370-2693(76)90417-2}{{\em Phys. Lett.}
  {\bfseries B60} (1976) 177}.
%%CITATION = PHLTA,B60,177;%%.

\bibitem{Kojima:2011ad}
K.~Kojima, K.~Takenaga, and T.~Yamashita, ``{Grand Gauge-Higgs Unification},''
  \href{http://dx.doi.org/10.1103/PhysRevD.84.051701}{{\em Phys. Rev.}
  {\bfseries D84} (2011) 051701},
\href{http://arxiv.org/abs/1103.1234}{{\ttfamily arXiv:1103.1234 [hep-ph]}}.
%%CITATION = ARXIV:1103.1234;%%.

\bibitem{Kojima:2016fvv}
K.~Kojima, K.~Takenaga, and T.~Yamashita, ``{Gauge Symmetry Breaking Patterns
  in an SU(5) Grand Gauge-Higgs Unification Model},''
  \href{http://dx.doi.org/10.1103/PhysRevD.95.015021}{{\em Phys. Rev.}
  {\bfseries D95} no.~1, (2017) 015021},
\href{http://arxiv.org/abs/1608.05496}{{\ttfamily arXiv:1608.05496 [hep-ph]}}.
%%CITATION = ARXIV:1608.05496;%%.

\bibitem{Burdman:2002se}
G.~Burdman and Y.~Nomura, ``{Unification of Higgs and Gauge Fields in
  Five-Dimensions},''
  \href{http://dx.doi.org/10.1016/S0550-3213(03)00088-9}{{\em Nucl. Phys.}
  {\bfseries B656} (2003) 3--22},
\href{http://arxiv.org/abs/hep-ph/0210257}{{\ttfamily arXiv:hep-ph/0210257
  [hep-ph]}}.
%%CITATION = HEP-PH/0210257;%%.

\bibitem{Lim:2007jv}
C.~Lim and N.~Maru, ``{Towards a Realistic Grand Gauge-Higgs Unification},''
  \href{http://dx.doi.org/10.1016/j.physletb.2007.07.053}{{\em Phys.Lett.}
  {\bfseries B653} (2007) 320--324},
\href{http://arxiv.org/abs/0706.1397}{{\ttfamily arXiv:0706.1397 [hep-ph]}}.
%%CITATION = ARXIV:0706.1397;%%.

\bibitem{Kojima:2017qbt}
K.~Kojima, K.~Takenaga, and T.~Yamashita, ``{The Standard Model Gauge Symmetry
  from Higher-Rank Unified Groups in Grand Gauge-Higgs Unification Models},''
\href{http://arxiv.org/abs/1704.04840}{{\ttfamily arXiv:1704.04840 [hep-ph]}}.
%%CITATION = ARXIV:1704.04840;%%.

\bibitem{Kim:2002im}
H.~D. Kim and S.~Raby, ``{Unification in 5-D SO(10)},''
  \href{http://dx.doi.org/10.1088/1126-6708/2003/01/056}{{\em JHEP} {\bfseries
  01} (2003) 056},
\href{http://arxiv.org/abs/hep-ph/0212348}{{\ttfamily arXiv:hep-ph/0212348
  [hep-ph]}}.
%%CITATION = HEP-PH/0212348;%%.

\bibitem{Fukuyama:2008pw}
T.~Fukuyama and N.~Okada, ``{A Simple SO(10) GUT in Five Dimensions},''
  \href{http://dx.doi.org/10.1103/PhysRevD.78.015005}{{\em Phys. Rev.}
  {\bfseries D78} (2008) 015005},
\href{http://arxiv.org/abs/0803.1758}{{\ttfamily arXiv:0803.1758 [hep-ph]}}.
%%CITATION = ARXIV:0803.1758;%%.

\bibitem{Hosotani:2015hoa}
Y.~Hosotani and N.~Yamatsu, ``{Gauge-Higgs Grand Unification},''
  \href{http://dx.doi.org/10.1093/ptep/ptv153}{{\em Prog. Theor. Exp. Phys.}
  {\bfseries 2015} (2015) 111B01},
\href{http://arxiv.org/abs/1504.03817}{{\ttfamily arXiv:1504.03817 [hep-ph]}}.
%%CITATION = ARXIV:1504.03817;%%.

\bibitem{Hosotani:2015wmb}
Y.~Hosotani and N.~Yamatsu, ``{Gauge-Higgs Grand Unification},'' {\em PoS}
  {\bfseries PLANCK2015} (2015) 058,
\href{http://arxiv.org/abs/1511.01674}{{\ttfamily arXiv:1511.01674 [hep-ph]}}.
%%CITATION = ARXIV:1511.01674;%%.

\bibitem{Yamatsu:2015rge}
N.~Yamatsu, ``{Gauge Coupling Unification in Gauge-Higgs Grand Unification},''
  \href{http://dx.doi.org/10.1093/ptep/ptw023}{{\em Prog. Theor. Exp. Phys.}
  {\bfseries 2016} (2016) 043B02},
\href{http://arxiv.org/abs/1512.05559}{{\ttfamily arXiv:1512.05559 [hep-ph]}}.
%%CITATION = ARXIV:1512.05559;%%.

\bibitem{Furui:2016owe}
A.~Furui, Y.~Hosotani, and N.~Yamatsu, ``{Toward Realistic Gauge-Higgs Grand
  Unification},'' \href{http://dx.doi.org/10.1093/ptep/ptw116}{{\em Prog.
  Theor. Exp. Phys.} {\bfseries 2016} (2016) 093B01},
\href{http://arxiv.org/abs/1606.07222}{{\ttfamily arXiv:1606.07222 [hep-ph]}}.
%%CITATION = ARXIV:1606.07222;%%.

\bibitem{Hosotani:2016njs}
Y.~Hosotani, ``{Gauge-Higgs EW and Grand Unification},''
  \href{http://dx.doi.org/10.1142/S0217751X16300313}{{\em Int. J. Mod. Phys.}
  {\bfseries A31} no.~20n21, (2016) 1630031},
\href{http://arxiv.org/abs/1606.08108}{{\ttfamily arXiv:1606.08108 [hep-ph]}}.
%%CITATION = ARXIV:1606.08108;%%.

\bibitem{Kawamura:2013rj}
Y.~Kawamura and T.~Miura, ``{Classification of Standard Model Particles in
  $E_6$ Orbifold Grand Unified Theories},''
  \href{http://dx.doi.org/10.1142/S0217751X13500553}{{\em Int. J. Mod. Phys.}
  {\bfseries A28} (2013) 1350055},
\href{http://arxiv.org/abs/1301.7469}{{\ttfamily arXiv:1301.7469 [hep-ph]}}.
%%CITATION = ARXIV:1301.7469;%%.

\bibitem{Hosotani:2017ghg}
Y.~Hosotani and N.~Yamatsu. in preperation.

\bibitem{Dynkin:1957ek}
E.~Dynkin, ``{Maximal Subgroups of the Classical Groups},'' {\em
  Trans.Am.Math.Soc.} {\bfseries 6} (1957) 245.

\bibitem{Dynkin:1957um}
E.~Dynkin, ``{Semisimple Subalgebras of Semisimple Lie Algebras},''
{\em Trans.Am.Math.Soc.} {\bfseries 6} (1957) 111.
%%CITATION = TAMTA,6,111;%%.

\bibitem{Cahn:1985wk}
R.~Cahn, {\em {Semi-Simple Lie Algebras and Their Representations}}.
\newblock Benjamin-Cummings Publishing Company,
1985.
\newblock
%%CITATION = INSPIRE-220438;%%.

\bibitem{Mckay:1977}
W.~Mckay, J.~Patera, and D.~Sankoff, {\em {The Computation of Branching Rules
  for Representations of Semisimple Lie Algebras}}.
\newblock New York Academic Press, 1977.
\newblock in ``Computers in Nonassociative Rings and Algebras'' edited by
  R.~E.~Beck and B.~Kolman.

\bibitem{McKay:1981}
W.~G. McKay and J.~Patera, {\em Tables of Dimensions, Indices, and Branching
  Rules for Representations of Simple Lie Algebras}.
\newblock Marcel Dekker, Inc., New York, 1981.

\bibitem{Fonseca:2011sy}
R.~M. Fonseca, ``{Calculating the Renormalisation Group Equations of a SUSY
  Model with Susyno},'' \href{http://dx.doi.org/10.1016/j.cpc.2012.05.017}{{\em
  Comput.Phys.Commun.} {\bfseries 183} (2012) 2298--2306},
\href{http://arxiv.org/abs/1106.5016}{{\ttfamily arXiv:1106.5016 [hep-ph]}}.
%%CITATION = ARXIV:1106.5016;%%.

\bibitem{Feger:2012bs}
R.~Feger and T.~W. Kephart, ``{LieART - A Mathematica Application for Lie
  Algebras and Representation Theory},''
  \href{http://dx.doi.org/10.1016/j.cpc.2014.12.023}{{\em Comput.Phys.Commun.}
  {\bfseries 192} (2015) 166--195},
\href{http://arxiv.org/abs/1206.6379}{{\ttfamily arXiv:1206.6379 [math-ph]}}.
%%CITATION = ARXIV:1206.6379;%%.

\bibitem{Kawamura:1999nj}
Y.~Kawamura, ``{Gauge Symmetry Breaking from Extra Space $S^1/Z_2$},''
  \href{http://dx.doi.org/10.1143/PTP.103.613}{{\em Prog. Theor. Phys.}
  {\bfseries 103} (2000) 613--619},
\href{http://arxiv.org/abs/hep-ph/9902423}{{\ttfamily arXiv:hep-ph/9902423
  [hep-ph]}}.
%%CITATION = HEP-PH/9902423;%%.

\bibitem{Kawamura:2000ev}
Y.~Kawamura, ``{Triplet-Doublet Splitting, Proton Stability and Extra
  Dimension},'' \href{http://dx.doi.org/10.1143/PTP.105.999}{{\em Prog. Theor.
  Phys.} {\bfseries 105} (2001) 999--1006},
\href{http://arxiv.org/abs/hep-ph/0012125}{{\ttfamily arXiv:hep-ph/0012125}}.
%%CITATION = HEP-PH/0012125;%%.

\bibitem{Hosotani:1983xw}
Y.~Hosotani, ``{Dynamical Mass Generation by Compact Extra Dimensions},''
\href{http://dx.doi.org/10.1016/0370-2693(83)90170-3}{{\em Phys.Lett.}
  {\bfseries B126} (1983) 309}.
%%CITATION = PHLTA,B126,309;%%.

\bibitem{Hosotani:1988bm}
Y.~Hosotani, ``{Dynamics of Nonintegrable Phases and Gauge Symmetry
  Breaking},''
\href{http://dx.doi.org/10.1016/0003-4916(89)90015-8}{{\em Annals Phys.}
  {\bfseries 190} (1989) 233}.
%%CITATION = APNYA,190,233;%%.

\bibitem{Randall:1999ee}
L.~Randall and R.~Sundrum, ``{A Large Mass Hierarchy from a Small Extra
  Dimension},'' \href{http://dx.doi.org/10.1103/PhysRevLett.83.3370}{{\em Phys.
  Rev. Lett.} {\bfseries 83} (1999) 3370--3373},
\href{http://arxiv.org/abs/hep-ph/9905221}{{\ttfamily arXiv:hep-ph/9905221}}.
%%CITATION = HEP-PH/9905221;%%.

\bibitem{Hosotani:2017krs}
Y.~Hosotani, ``{New Dimensions from Gauge-Higgs Unification},''
\href{http://arxiv.org/abs/1702.08161}{{\ttfamily arXiv:1702.08161 [hep-ph]}}.
%%CITATION = ARXIV:1702.08161;%%.

\bibitem{Banks1976}
J.~Banks and H.~Georgi, ``{Comment on Gauge Theories Without Anomalies},''
\href{http://dx.doi.org/10.1103/PhysRevD.14.1159}{{\em Phys. Rev.} {\bfseries
  D14} (1976) 1159--1160}.
%%CITATION = PHRVA,D14,1159;%%.

\bibitem{Dienes:1996yh}
K.~R. Dienes and J.~March-Russell, ``{Realizing Higher Level Gauge Symmetries
  in String Theory: New Embeddings for String GUTs},''
  \href{http://dx.doi.org/10.1016/0550-3213(96)00406-3}{{\em Nucl. Phys.}
  {\bfseries B479} (1996) 113--172},
\href{http://arxiv.org/abs/hep-th/9604112}{{\ttfamily arXiv:hep-th/9604112
  [hep-th]}}.
%%CITATION = HEP-TH/9604112;%%.

\bibitem{Dienes:1996du}
K.~R. Dienes, ``{String Theory and the Path to Unification: A Review of Recent
  Developments},'' \href{http://dx.doi.org/10.1016/S0370-1573(97)00009-4}{{\em
  Phys. Rept.} {\bfseries 287} (1997) 447--525},
\href{http://arxiv.org/abs/hep-th/9602045}{{\ttfamily arXiv:hep-th/9602045
  [hep-th]}}.
%%CITATION = HEP-TH/9602045;%%.

\end{thebibliography}\endgroup

\end{document}